\documentclass{article}

\usepackage{graphicx}
\usepackage{subfigure}
\usepackage{color}
\usepackage{natbib}
\usepackage{amsmath}
\usepackage{subeqnarray}

\renewcommand{\vec}{\mathbf}

\newsavebox{\astrutbox}
\sbox{\astrutbox}{\rule[-5pt]{0pt}{20pt}}

\begin{document}

\title{Hydrodynamics of flagellated microswimmers near free-slip interfaces}

\author{D. Pimponi$^1$
M. Chinappi$^2$
P. Gualtieri $^1$\break
and C. M.  Casciola$^1$}%

\date{}

\maketitle

\begin{abstract}
The hydrodynamics of a flagellated microorganism is investigated 
when swimming close to a planar free-slip surface by means of numerical
solutions of the Stokes equations obtained via a Boundary Element Method. 
Depending on the initial condition, the swimmer can either escape from the 
free-slip surface or collide with the boundary. Interestingly,  the microorganism does 
not exhibit a stable orbit.  
Independently of escape or attraction to the interface, close to a free-slip surface,
 the swimmer  follows a counter-clockwise trajectory, in agreement with experimental findings,~\cite{di2011swimming}. 
The hydrodynamics is indeed modified by the
free-surface. In fact, when the same swimmer moves close to a no-slip wall,  a set of 
initial conditions exists which result in stable orbits. 
Moreover when moving close to a free-slip or a no-slip boundary the swimmer
assumes a different orientation with respect to its trajectory.
Taken together, these results contribute to shed light on the hydrodynamical
behaviour of microorganisms close to liquid-air interfaces which are relevant for the formation of 
interfacial biofilms  of aerobic bacteria.
\end{abstract}


\section{Introduction}\label{sec:intro}

The recent development of micro-fluidic based devices has pushed the
scientific community to reconsider classical issues in fluid dynamics
such as the hydrodynamical behavior of micro-scale bodies
in the creeping regime. For instance, propulsion of microswimmers,
see e.g.~\cite{fauci2006biofluidmechanics,lauga2009hydrodynamics,guasto2012fluid,elgeti2014physics}, 
has been addressed for the production of energy oriented to 
micro-devices~\citep{di2010bacterial,sokolov2010swimming} 
and for the self-propulsion of micro-robots~\citep{pak2012micropropulsion,alouges2013optimally}. 
The microswimmer's behavior in presence of confining surfaces is also 
crucial for biofilm formation. Most studies~\citep{pratt1998genetic, rusconi2010laminar}
focused on the biofilm formation at  solid-liquid interfaces.
Biofilm formation near air-liquid interfaces is also a problem of significant concern, since
the liquid-air biofilm can be  advantageous for aerobic microorganisms, 
providing them access to oxygen (from the air) and nutrients 
(from the liquid) at the same time~\citep{constantin2009bacterial, koza2009characterization}.
Several species of flagellated bacteria are known to form biofilms at liquid/air interfaces, e.g. 
Bacillus subtilis, Bacillus cereus and Pseudomonas fluorescens. 
For instance, the preparation of the well known Natto from Japanese cooking, consisting of fermented soy beens, involves 
B. subtilis in significant concentrations~\citep{chantawannakul2002characterization}. In contrast, some strains  of B. cereus are known to be harmful 
to humans and cause foodborne illness~\citep{ehling2004bacillus}. P. fluorescens, instead, is able to contaminate heparinized saline flushes used in cancer therapy~\citep{gershman2008multistate}.  Despite its biological implications, the mechanics  of flagellated swimmers close to  a liquid-air interface is much less explored than the case of a liquid-solid interface.

At micro-scales viscous effects overwhelm inertia leading to the development
of apparently counterintuitive swimming strategies as proved by the 
Scallop theorem~\citep{purcell1977life,lauga2011life}. In a nutshell, 
if the swimmer deforms through a sequence of body configurations which are
periodic and time-reversible (reciprocal motion), its average motion is zero. 
The reciprocal motion can be exploited for locomotion only when 
non-linear or memory effects are relevant.
Examples are the non-Newtonian behavior of
the fluid~\citep{qiu2014swimming,lauga2009life,keim2012fluid} or motions occurring close to 
deformable interfaces~\citep{trouilloud2008soft}. Microorganisms developed 
many different strategies based on nonreciprocal effects to overcome the above 
restrictions. The \textit{Spiroplasma} deforms its cytoskeleton by propagating 
pairs of kinks~\citep{trachtenberg2003bacterial,yang2009kinematics}.
Other microswimmers exploit their cilia wavy motion~\citep{maxey2011biomimetics}
as done by \textit{Paramecium}~\citep{jana2012paramecium}.

However, many microorganisms take advantage of single or multiple 
flagella, such as the \textit{Caulobacter crescentus}, which has a single 
(right-handed) helical filament driven by a rotary motor~\citep{li2006low} and the 
\textit{Escherichia coli} that has multiple flagella~\citep{berg2004coli}. 
The flagellar motor which activates and controls the filament rotation is 
able to switch between both rotation directions~\citep{wang2014switching}, 
and, as a first approximation, the torque applied to the filament is 
constant~\citep{berg2004coli}. 
In the case of the
\textit{Escherichia coli} the flagella arrange in a bundle, characterized by 
 a left-handed rotation of the motor
which 
confers to bacteria a smooth forward motion (\textit{run} phase).
The flagella is also able to invert the rotation direction in 
order to let the bacteria at rest and change its direction 
(\textit{tumble} phase).	
Due to their internal structure, the filaments can only assume twelve prescribed 
shapes (helical polymorphic states) but only one of those is the 
\lq\lq normal state\rq\rq, i.e. it is the most observed one during the \textit{run} 
phase~\citep{darnton2007force,vogel2010force}.
Hence, the flagella can be 
modeled as a rigid single filament~\citep{phan1987boundary,shum2010modelling} 
rotating around the bacterial head axis.

The main purpose of this work is an extensive analysis of the hydrodynamical
behavior of a flagellated microswimmer close to air-liquid interfaces, with particular 
attention to free surfaces. 
For the sake of definiteness, we  focus on a simplified geometry modeling the configuration of E. coli whose swimming  is probably the most widely studied  from both experimental and numerical point of view. Anyhow, in order to explore the effect of different geometrical parameters, several modified configurations are also discussed.

Flagellated swimmers close to a free-surface have been already addressed by simplified models, see e.g. 
~\cite{crowdy2011two, di2011swimming, lopez2014dynamics} where two-dimensionality, resistive force theory, and multipole expansions techniques were exploited, respectively.
To the best of our knowledge, the present work  addresses the first fully three dimensional numerical simulation of a swimmer in presence of a liquid-air interface.
Given the linearity of
the Stokes equations which are the appropriate model for creeping flows, the numerical
approach exploits the Boundary Element Method (BEM). The BEM 
can easily handle the complex geometry of the microswimmer and account for
different boundary conditions. Moreover, a flat and 
infinitely extended free surface and/or solid wall (when needed for comparisons) 
can be easily included by considering the appropriate Green's function, thus 
avoiding any undesired effect of the numerical truncation of the domain.

There is a wide body of literature available dealing with the motion of 
microorganisms in free space or close to solid boundaries. 
Since the first studies by~\cite{taylor1951analysis}, the 
attention was principally focused on microorganisms whose 
flagellum was modeled as an infinite cylindrical filament
in an unbounded fluid endowed with small amplitude wavy motion. 
In fact, it was initially believed that the flagella were only moved by  
wave propagation. However~\cite{berg1973bacteria} showed that bacteria could
also rotate their flagella in a corkscrew-like motion, moving the flagellar bundle as 
a single filament.
The flagellum has a large aspect ratio, with length exceeding 
thickness by even more than two orders of magnitude~\citep{lauga2009hydrodynamics}. 
This particular geometry pushed the adoption of a 
Slender Body Theory (SBT) that has been exploited in several studies to evaluate 
the translational velocity and/or the torque applied by the 
swimmer~\citep{hancock1953self}. 
Successively \cite{higdon1979hydrodynamic,higdon1979hydrodynamics} 
transformed the Stokes equations into a system of singular integral equations 
accounting for the swimmer translational and angular velocities. He added to the SBT 
 a variable strength of the singularity along the flagellum centerline, 
thus modelling different centerline geometries. In this case both the planar 
sinusoidal motion and the rotation about the body axis were amenable to modelling.
By an appropriate system of images the SBT could also account for a spherical cell body and 
the presence of a wall. However too many restrictions still confined the SBT application 
to extremely simplified configurations. 

In recent years the increase of computational resources, led to a more 
extensive use of the BEM which overcomes many drawbacks of the SBT in dealing 
with microswimmers both in free space and confined conditions. \cite{phan1987boundary} 
used the BEM to study the motion of a microorganism in free space. 
Successively,~\cite{ramia1993role} applied the BEM 
to the interaction between the swimmer and a solid wall, showing that, 
when swimming close to a solid wall, the swimmer exhibited a 
circular motion. The BEM was also 
used to study the interactions between two neighboring flagellated microswimmers
highlighting the possibility of a coordination between their flagellar motion 
in order to maximize their velocities~\citep{ramia1993role}. 
 
More recently,~\cite{lauga2006swimming} experimentally investigated the motion 
of an E. coli near a solid wall and found, as predicted by~\cite{ramia1993role}, 
a circular clockwise motion.  In the paper the authors also
provided a simple theoretical model which was able to explain their experimental
observations. Interestingly, the same model suggested that the bacteria
should reverse its rotation when swimming in proximity of a free surface. 
The same authors also showed that the swimmer is attracted by a solid wall.
This behavior was deeply investigated a few years later 
by~\cite{giacche2010hydrodynamic} in a numerical work which highlighted 
how the bacteria could move at a stable distance from a wall. At the same 
time~\cite{shum2010modelling} investigated the motion of a microswimmer 
close to solid walls by considering many geometrical configurations
and relating the shape of the body to the propulsion efficiency and to the
possibility of achieving different motions, i.e. to be attracted by the wall, 
to escape from the wall or to reach stable circular orbits at a given 
wall-normal distance.
The same authors~\citep{shum2012effects} also focused on the flexible hooks 
linking each flagellum to the cell. They studied the modifications in the microswimmer trajectories
when changing the hook rigidities. They found that, within an intermediate range of rigidities, the 
swimmer behavior doesn't change too much with respect to the simpler model of 
rigid hook. The same work highlighted how, for particular values of relative hook stiffness, 
there is a transient phase of periodic motion with constant average distance from the wall, 
leading to boundary accumulation. 
The tendency of swimming microorganisms to accumulate near solid 
walls~\citep{li2009accumulation}, with particular emphasis on collisions with the surface 
and rotational Brownian motion, has also been investigated and linked 
to the swimming speed and the cell size~\citep{li2011accumulation} of the microswimmer.
Recent works extended the investigation about the motion of microswimmers 
to more complex surfaces: a clean fluid-fluid interface, a slipping rigid wall, and a fluid interface covered 
by surfactants~\citep{lopez2014dynamics}.  
The authors used an asymptotic, far field approximation 
to represent the actual swimmer, 
retaining
information about velocities and rotations.
The case of two fluid interfaces, in the limit of vanishing 
viscosity of one of them, allowed to describe a free surface. 
In such conditions, the swimmer exhibited counter-clockwise motion.
This confirmed the results anticipated by~\cite{lauga2006swimming}. 
Also~\cite{di2011swimming} supported by means of experimental 
observations the theoretical prediction made through a simplified model based 
on the method of images and the resistive force theory. Even if neglecting all the 
dynamics of the swimmer outside the interface plane, this work provided a simple 
explanation for a counter-clockwise motion over a perfectly slipping surface, showing 
a good agreement between the results of the simplified model and the experimental 
observation. 
Briefly, ~\cite{lauga2006swimming} explain that, when swimming above a no slip surface, a positive rotation rate of the swimmer head around its longitudinal axis produces a lateral force which is opposite to the one induced by the negative tail rotation rate. As a consequence, a net torque 
normal to the wall is exerted on the body such that the swimmer follows a clockwise (CW) trajectory.
Otherwise, when swimming close to a liquid-air interface,~\cite{di2011swimming} devise a simplified model based on resistive force theory endowed with suitable symmetries to satisfy the free slip condition at the interface. 
In this model, the swimmer moves under the effect of the velocity generated by its mirror image below the interface.
The counter-rotating image head produces on the swimmer head a lateral velocity which is opposite to the force that is exerted in the case of a solid wall.
Such relative velocity gives rise to a corresponding viscous force in the same direction. Since the same reasoning applies to the counter rotating tail, the overall torque on the microswimmer  is also opposite to the one experienced on a no-slip wall, hence
 a net CCW motion is produced. 
In literature a clockwise motion 
has been observed~\citep{lemelle2010counterclockwise} 
also in presence of a free surface. 
Based on experimental observations,~\cite{morse2013molecular} attributed such a behavior to 
the molecular adsorption (due to the presence of biological material in the growth medium) 
altering the rheological properties of the air/water interface, thus determining the swimming 
pattern of nearby cells. 

In principle, the mechanical causes that can affect the swimming direction are: 
$i)$ the rotation direction of the flagellum bundle;
$ii)$ the effective boundary condition at the planar surface.
This study focuses on the point $ii)$ through the numerical simulation of
the motion of a \textit{E. coli}-like microswimmer   
close to free-slip and no-slip surfaces, assuming a standard left-handed arrangement for 
the flagellum bundle. 
The aim is investigating the behavior of the swimmer by 
addressing in full detail the complete three dimensional nature of the hydrodynamical 
interaction between the swimmer and the surface. 
It is worth noting that free-slip and no-slip are the limits 
of the more general Navier boundary condition that 
connects the velocity at the liquid boundary with the tangential 
stress~\citep{bazant2008tensorial}.
For liquid water moving on a solid surface, the actual slip is 
negligible at micro-scale also for hydrophobic coatings 
\citep{chinappi2010intrinsic,sega2013regularization} and only the presence 
of vapor bubbles trapped in the surface asperities (superhydrophobic surfaces) 
leads to significant slippage~\citep{gentili2014pressure,bolognesi2013novel}, potentially
altering the motion of particles close to the surface
~\citep{pimponi2014mobility,nizkaya2014flows}. Hence, the no-slip boundary is, for
the present purposes, a reliable model of a rigid wall. 
On the other hand, the proper boundary conditions at the liquid-air interface 
are impermeability and continuity of the tangential stress 
components, that reduces to free-slip (zero tangential stress) since 
air density is order of magnitude smaller than the liquid one.

The paper is organized as follows:
the geometrical model of the swimmer and the BEM is addressed in 
\S~\ref{sec:model}. \S~\ref{sec:FreeSurf} reports the salient results 
concerning the motion of the swimmer in presence of a free surface.
Finally, we discuss the major findings and point out the main conclusions 
(\S~\ref{sec:conclusions}), giving a perspective view of future work. 
\section{Boundary integral formulation for the microswimmer}\label{sec:model}

\subsection{Swimmer model}\label{ssec:swim_mod}

This section concerns the modelling of a microswimmer 
inspired to \textit{Escherichia coli}. This bacteria has 
been deeply investigated in the literature and a lot of  information about 
its geometry and propulsion mechanism is available. \textit{E. coli} has a 
cell length which varies between $1.6$ and $3.9 \, \mu m$, 
the cell width ranges between $0.9$ and $1.7 \, \mu m$, 
with a resulting cell volume from 
$1.5$ to $4.4 \, \mu m^3$~\citep{volkmer2011condition} and an
average length of flagella of about $7 \mu m$~\citep{lauga2006swimming}.

\begin{figure}
  \centerline{\includegraphics[height=6cm,width=12cm]{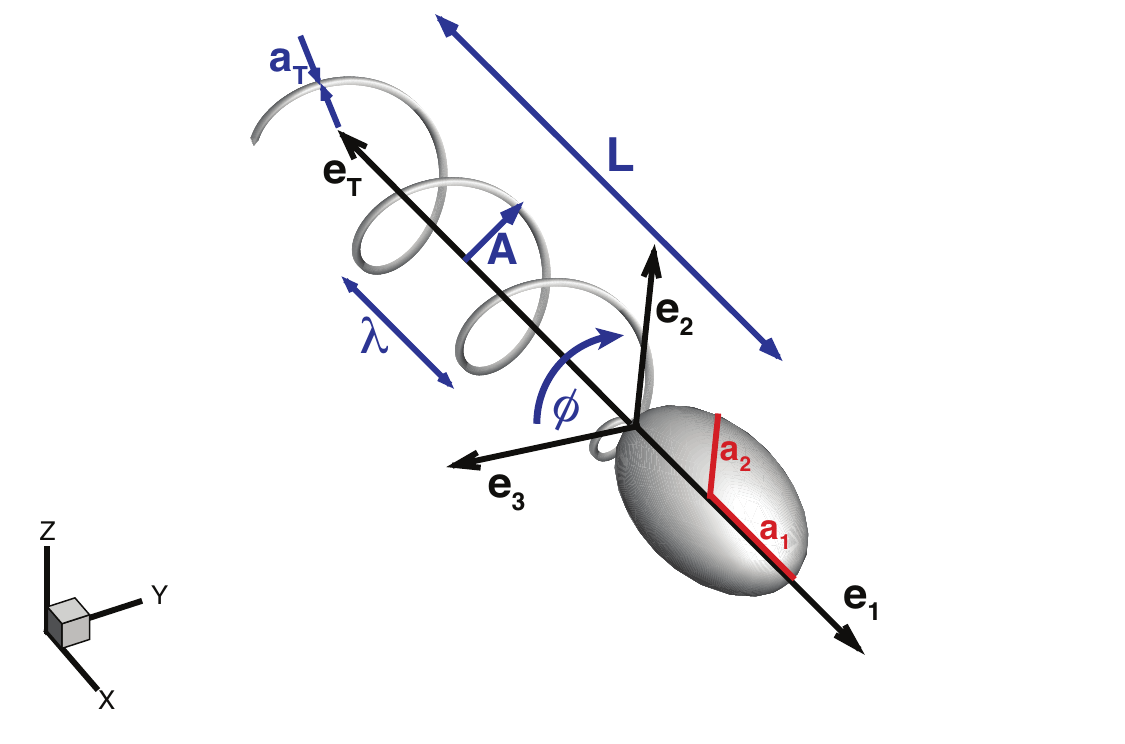}}
  \caption{The model of the flagellated swimmer comprises an
           ellipsoidal head (cell) and a tubular, helical, rigid, tail (flagellum).
           The tail rotates about its axis $\vec{e}_T$ of an angle $\phi(t)$ . 
           The dimensionless semi-axis of the head are $a_1$ and $a_2$. 
           The head aspect ratio is kept constant, $AR=a_1/a_2=2$, and the 
           equivalent radius (i.e. the radius $\bar{a}$ of the sphere with same 
           volume) is the assumed reference length.
           The dimensionless tail length is $L=7$, with a cross-section 
           radius $a_t=0.05$. The dimensionless helix amplitude and 
           wavelength are $A$ and $\lambda$, respectively. 
           $\{ \vec{e_1},\vec{e_2},\vec{e_3} \}$ are the orthonormal 
           vectors of the frame attached to the swimmer head (body frame). 
           $\vec{e_1}$ is longitudinal and identifies the swimmer 
           orientation with respect to the unit vectors of the 
           fixed frame $\{ \vec{X},\vec{Y},\vec{Z} \}$.
  }
\label{fig:geometry}
\end{figure}
Figure~\ref{fig:geometry} sketches the simplified geometry comprising the 
ellipsoidal axisymmetric cell and the corkscrew tail. Hereafter the equivalent 
radius of the ellipsoidal cell $\bar{a}$ (the radius of the sphere having the same 
volume) will be used as reference length-scale, i.e. the dimensionless cell volume is 
$V = 3 V'/ {4 \pi \bar a}^3 = 1$, where the prime identifies dimensional 
quantities. The cell aspect ratio is $AR=a_1/a_2=2$, being $2 a_1$ and $2 a_2$ the 
longitudinal and the transversal (dimensionless) axis. 
Following~\cite{shum2010modelling}, the tail bundle is modelled 
as a single helix with radius $a_T=0.05$. The tail 
rotates around its axis $\vec{e_T}$ (see figure~\ref{fig:geometry}). 
The dimensionless axial length of the helix is 
$L$, $A$ denotes the helix amplitude and $\lambda$ its wavelength. 
In the present work we selected typical values 
for the tail length and cell axes, namely $L = 7$, $a_1 = 1.6$ and $a_2 = 0.8$, 
while the amplitude $A$ and the wavelength $\lambda$ of the tail are systematically 
varied. With the above choices, the
swimmer is rescaled into an actual \textit{E. coli} 
by assuming $\bar{a}=1 \,\, \mu m$.
\begin{figure}
  \centerline{\includegraphics[height=6cm,width=12cm]{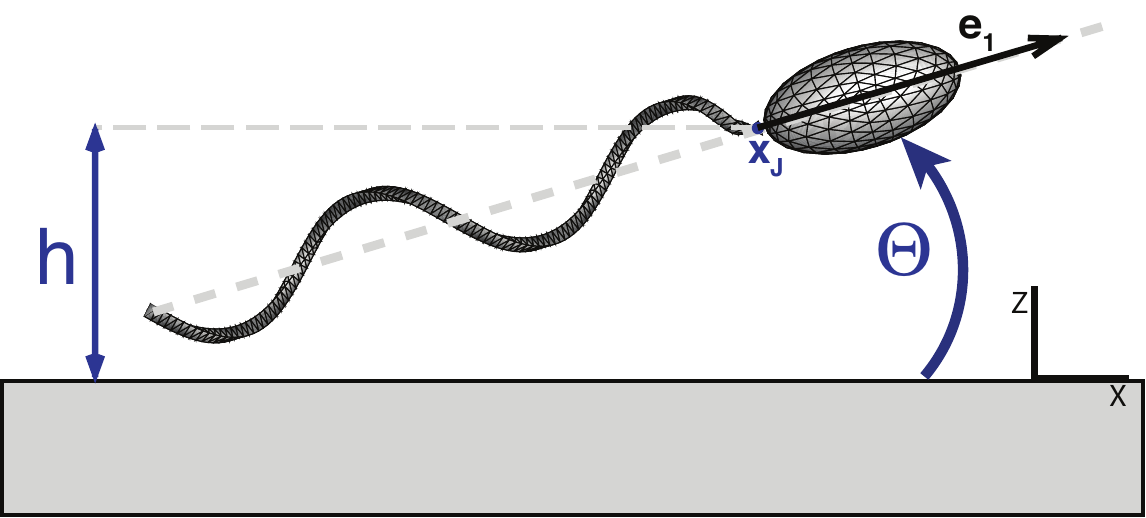}}
  \caption{Sketch of the discretized swimmer near a planar surface.
           The configuration is identified by three parameters,
           namely, the distance $h$ of the reference point $\vec{x}_J$ 
           from the plane and the pitch angle $\Theta$. 
           The third parameter is the tail rotation angle $\phi$ defined in
           figure~\ref{fig:geometry}. 
           }
\label{fig:pitch_angle}
\end{figure}

It is useful to introduce a body reference frame with orthonormal base vectors 
$\{ \vec{e_1},\vec{e_2},\vec{e_3} \}$ 
where $\vec{e_1}=-\vec{e}_T$ is longitudinal (see figure~\ref{fig:geometry}).
The subscript $H$ identifies the head and the subscript $T$ refers 
to the tail, which can rotate with respect to the head about its axis $\vec{e_T}$. 
During the tail rotation around $\vec{e}_T$, each point of the rigid tail  
describes a circle in the plane normal to $\vec{e}_T$, being $\phi$ the 
corresponding rotation angle (or flagellum phase).
The time derivative $\dot{\phi}(t)$ is the tail rotational velocity $\Omega_T$. 
The swimmer position is identified in a fixed reference frame 
with base 
$\{\vec{X},\vec{Y},\vec{Z}\}$ by the three coordinates of the cell-to-tail 
junction point $\vec{x}_J$, see figure~\ref{fig:pitch_angle}.
The swimmer translates with dimensionless velocity $\vec{U}=\vec{U'}/v$
and has angular velocity $\vec{\Omega}_H=\vec{\Omega'}_H \bar{a}/ v$,
where $v \simeq 20\mu m /s$ is a typical swimming velocity 
taken as reference quantity.
The $\{ \vec{X},\vec{Y} \}$ coordinate plane 
is taken to coincide with the planar boundary, either
the solid wall or the free surface, see figure~\ref{fig:pitch_angle}.

The kinematics of the swimmer is described by seven degrees of freedom, 
the junction position $\vec{x}_J(t)$, the flagellum phase $\phi(t)$ and the 
three parameters defining the rotation of 
the head. We stress that the discretely evolved rotation matrix should
belong to the proper matrix subspace, namely the $\rm SO(3)$ subgroup. 
This is enforced through a description in terms 
of quaternions with unit norm, see e.g.~\cite{diebel2006representing}. 
It follows,
\begin{eqnarray}
\vec{\dot{x}}_J &=& \vec{U}(t) \,\, , \label{eqn:kinematic1} \\
\dot{\phi} &=& \Omega_T (t) \,\, , \label{eqn:kinematic2} \\
\vec{\dot{q}} &=& \frac{1}{2}\vec{S}(\vec{q}) \, \vec{\Omega}_H(t)\,\, , \label{eqn:kinematic3}
\end{eqnarray}
where $\vec{q}= (q_0 , q_1 , q_2 , q_3)$ is the quaternion vector
with $\vert \vec{q} \vert =1$. When $\vec{\Omega}_H(t)$  
is expressed in the 
$\{\vec{X},\vec{Y},\vec{Z}\}$ base, $\vec{S}$ reads 
\begin{equation}
  \vec{S}(\vec{q})=
  \left[ {\begin{array}{rrr}
   -q_1 & -q_2 & -q_3 \\
    q_0 & -q_3 &  q_2 \\
    q_3 &  q_0 & -q_1 \\
   -q_2 &  q_1 &  q_0 \\
  \end{array} } \right] \, .
\label{eqn:w_def}
\end{equation}
Given the quaternion $\vec{q}(t)$, the components of the body frame unit vectors
can be reconstructed as
\begin{eqnarray}
\nonumber 
\vec{e_1}= \left( q_0^2+q_1^2-q_2^2-q_3^2, 2 (q_1 q_2+q_0 q_3), 2 (q_1 q_3 - q_0 q_2) \right) \,\, , \\
\vec{e_2}= \left( 2 (q_1 q_2 - q_0 q_3), q_0^2-q_1^2+q_2^2-q_3^2, 2(q_2 q_3 + q_0 q_1)  \right) \,\, , \\
\nonumber 
\vec{e_3}= \left( 2(q_1 q_3+q_0 q_2), 2(q_2 q_3 - q_0 q_1), q_0^2-q_1^2-q_2^2+q_3^2 \right) \,\, . 
\label{eqn:bodyunitvectors}
\end{eqnarray}
Time integration of equations (\ref{eqn:kinematic1}),(\ref{eqn:kinematic2}),
(\ref{eqn:kinematic3}), allows  to track the swimmer trajectory once 
$\vec{U}$, $\vec{\Omega}_H$ and $\Omega_T$ are determined by the solution of the 
hydrodynamical interactions between the swimmer and the fluid, 
as discussed in the next section. Since the
numerical integration error in the equations for the quaternion may affect
its norm, the quaternion is normalized to unit length at each time step.
It easily shown that the accuracy 
of the integration scheme is exactly preserved by this procedure.

\subsection{Hydrodynamic model}\label{ssec:math_model}

Given the characteristic length scale $\bar{a}$ and swimming velocity $v$ of 
the microorganism, 
the typical Reynolds number in water is 
$Re=\rho \bar{a} v / \mu \simeq 2*10^{-5}$ where $\mu$ and $\rho$ 
are the water viscosity and density, respectively.
Hence the inertial terms are negligible and, for Newtonian fluids, the
flow in the domain ${\cal D}$ is described by the Stokes equations, 
\begin{subeqnarray}
\boldsymbol{\nabla} \cdot \vec{u} &=& 0 \,\, , \label{eqn:Continuity} \\
\nabla^2 \vec{u}-\boldsymbol{\nabla} p &=& 0 \,\, ,
\label{eqn:Stokes}
\end{subeqnarray}
where $\vec{u}=\vec{u'}/v$ and $p=p' \bar{a}/(\mu v)$.

The flow velocity $\vec{u}$ is forced by the tension 
$\vec{f}$ at the swimmer surface acting on the fluid.
The swimmer propulsion 
is due to the internal (constant) torque $\tau_M = \tau'_M/(\mu v \bar{a}^2)$
exchanged between head and tail.
Being a free body, the total force and torque on the swimmer (head and tail) 
is zero, 
\begin{subeqnarray} 
\int_{H \cup T}{\vec{f} dS} &=& 0 \,\, , \\
\int_{H \cup T}{\vec{r} \wedge \vec{f} dS}&=&0 \,\, .
\label{eqn:balances}
\end{subeqnarray}
where $\vec{r} = \vec{x} - \vec{x}_J$. 
The torque $\tau_M$ exerted on the tail is balanced by the torque 
produced by the fluid stresses $\vec{f}$ on the tail boundary, namely
\begin{equation}
\vec{e_T} \cdot \int_{T}{\vec{r} \wedge \vec{f} dS}=-\tau_M \,\, .
\label{eqn:Torque}
\end{equation}

The system ~(\ref{eqn:Stokes}),
~(\ref{eqn:balances}) and ~(\ref{eqn:Torque}) 
needs proper boundary conditions 
at the swimmer surface and external boundaries, i.e. wall or free surface.
On the microswimmer surface the no-slip 
condition yields
\begin{subeqnarray} 
\vec{u}(\vec{x}) &=& \vec{U}+\vec{\Omega}_H \wedge \vec{r}, \,\,\,\,\,\,\,\,\,\,\,\,\,\,\,\,\,\,\,\,\,\,\,\,\,\,\,\,\,\,\,\, (\vec{x}\in H) \,\,,  \\
\vec{u}(\vec{x}) &=& \vec{U}+(\vec{\Omega}_H + \Omega_T \vec{e}_T) \wedge \vec{r}, \,\,\,\,\, (\vec{x}\in T) \,\, .
\label{eqn:swim_velocity}
\end{subeqnarray}
No-slip boundary condition is also used to model the solid wall.
For the free surface vanishing 
normal fluid velocity (impermeability) and zero tangential stresses 
(free slip condition) are the appropriate prescriptions,
\begin{subeqnarray}
\vec{u} \cdot \vec{n} &=& 0 \,\, , \label{eqn:impermeab} \\
(\left( \vec{\nabla}\vec{u} + (\vec{\nabla} \vec{u})^\mathrm{T} \right) \cdot \vec{n})(\vec{I}-
\vec{n}\otimes\vec{n}) &=& 0 \,\, .
\label{eqn:noshear}
\end{subeqnarray}

\subsection{Boundary Element Method}\label{ssec:num_solution}

The solution of the Stokes equations~(\ref{eqn:Stokes}) can be 
recast in integral form
\begin{equation}
E(\vec{x}_0) \, u_i(\vec{x}_0)
=
\int_{\partial {\cal D}}
\left[ u_j(\vec{x}) T_{ikj}(\vec{x},\vec{x}_0) n_k(\vec{x}) - G_{ij}(\vec{x},\vec{x}_0) 
f_j(\vec{x}) \right] dS
\label{eqn:BIM}
\end{equation}
where $E(\vec{x_0})=1$ for points belonging to the interior 
of the domain. 
$G_{ij}(\vec{x},\vec{x}_0)$ is the 
free-space Green function, i.e. the $i-$th component of the
velocity at $\vec{x}$ induced by a Dirac $\delta$ like force
at $\vec{x}_0$ acting in direction $j$, and 
$T_{ikj}(\vec{x},\vec{x_0})$
the associated stress tensor~\citep{happel1983low}.
Equation (\ref{eqn:BIM}) expresses the $i$-th velocity component at 
the collocation point $\vec{x}_0$ in the fluid domain ${\cal D}$.
It requires the knowledge of the velocity $u_j(\vec{x})$ and the stresses $f_j(\vec{x})$ 
at the boundary $\partial {\cal D}$.
The integral representation (\ref{eqn:BIM}) can be turned 
into a boundary integral equation in the limit $\vec{x}_0 \to \partial {\cal D}$.
The resulting expression can be written in the same form (\ref{eqn:BIM}) now with
$E(\vec{x}_0) = 1/2$ and the integral understood in the Cauchy principal 
value sense.
We emphasize that the boundary $\partial {\cal D}$ 
consists of two disjointed parts: the swimmer surface and the planar interface.
The integral can be restricted to 
the swimmer's surface by exploiting the symmetries 
associated with the boundary condition at the planar interface. 
This is tantamount to 
the use of the appropriate  Green's function, see appendix~\ref{appA}, for the free-slip case 
and \cite{blake1971note} for the no-slip case.

Due to the boundary condition~(\ref{eqn:swim_velocity}) 
the fluid velocity at the swimmer surface is 
expressed in term of seven unknowns, namely, 
the velocity of the junction point $\vec{U}$, the angular velocity
$\vec{\Omega}_H$, and the tail rotation velocity $\Omega_T$,
which together with the tension $\vec{f}$ at 
the swimmer surface complete the set of unknowns. 
The system of equations consists of the vector boundary 
integral equation (\ref{eqn:BIM}), the global force and 
torque balance (\ref{eqn:balances}), and the 
torque balance for the swimmer tail (\ref{eqn:Torque}).
It can be shown that the solution exists and is unique
\citep{ladyzhenskaya1969mathematical}.
 
The above system is discretized by means of $N$ curved 6-point elements 
(typically $N = 3518$ total elements,  with $N_H =  512$ panels on the swimmer head), 
as sketched in figure~\ref{fig:pitch_angle}, with piecewise constant shape 
functions~\citep{pozrikidis1992boundary}.  By selecting the center of each element as 
a collocation point $\vec{x_0}$, the boundary integral equation~(\ref{eqn:BIM}) with 
$E(\vec{x_0})=1/2$ is recast in a system of $3N$ scalar equations for the stresses.
As will be shown in \S~\ref{sec:FreeSurf} the swimmer may 
approach the interface. In this case some of the panels belonging to the swimmer and to 
its image may get very close. This potentially spoils the accuracy of the integrals
providing the influence coefficients appearing in the discrete equations.
Care is taken in the simulations discussed below to have the diameter of 
the typical panel smaller than the distance between the body and its image
thereby preventing the undesired loss of accuracy.
Concerning the solution of the algebraic system, inverting a matrix coming from 
the discretization of boundary integral 
equations may require some care, since the algebraic system may be prone to severe 
ill-conditioning. An example is provided by the piecewise constant  approximation of 
a first kind Fredholm operator  \citep{hsiao1973solution}. 
In the context of the Stokes equations,  such operator arises when solving for 
the stresses given the velocity.  
However, the problem of the swimmer  addressed in the present paper 
does not involve  a pure first kind Fredholm operator, since the equations are augmented by  
the free-body constrains (zero net forces and torques) and by the equation enforcing the internal torque, with 
unknowns the stresses on the body plus the rigid body velocities and the relative  tail-body rotation rate. 
The spectral properties of the ensuing matrix differ substantially from those of a pure first kind Fredholm equation. 
In fact, the analysis based on the singular value decomposition, \citep{golub2012matrix}, excludes matrix ill-conditioning.
As a consequence,  the complete system of $3N+7$ algebraic equations for the  $3N+7$ unknowns collectively denoted  $\boldsymbol{\chi}$, 
$\vec{A} \cdot \boldsymbol{\chi} = \vec{b}$,  can be solved by standard techniques, like the Gauss-Jacobi method here implemented in 
an in-house MPI parallel  code.
Once the solution $\boldsymbol{\chi}$ is obtained, the integration of the kinematic  equations~(\ref{eqn:kinematic1}),~(\ref{eqn:kinematic2}),
~(\ref{eqn:kinematic3}),  performed via  a third order low-storage Runge-Kutta method, allows to track the swimmer trajectory.

\section{Swimming close to a free surface: results and discussion}\label{sec:FreeSurf}

 \begin{figure}
  \subfigure[]{\includegraphics[width=0.495\textwidth]{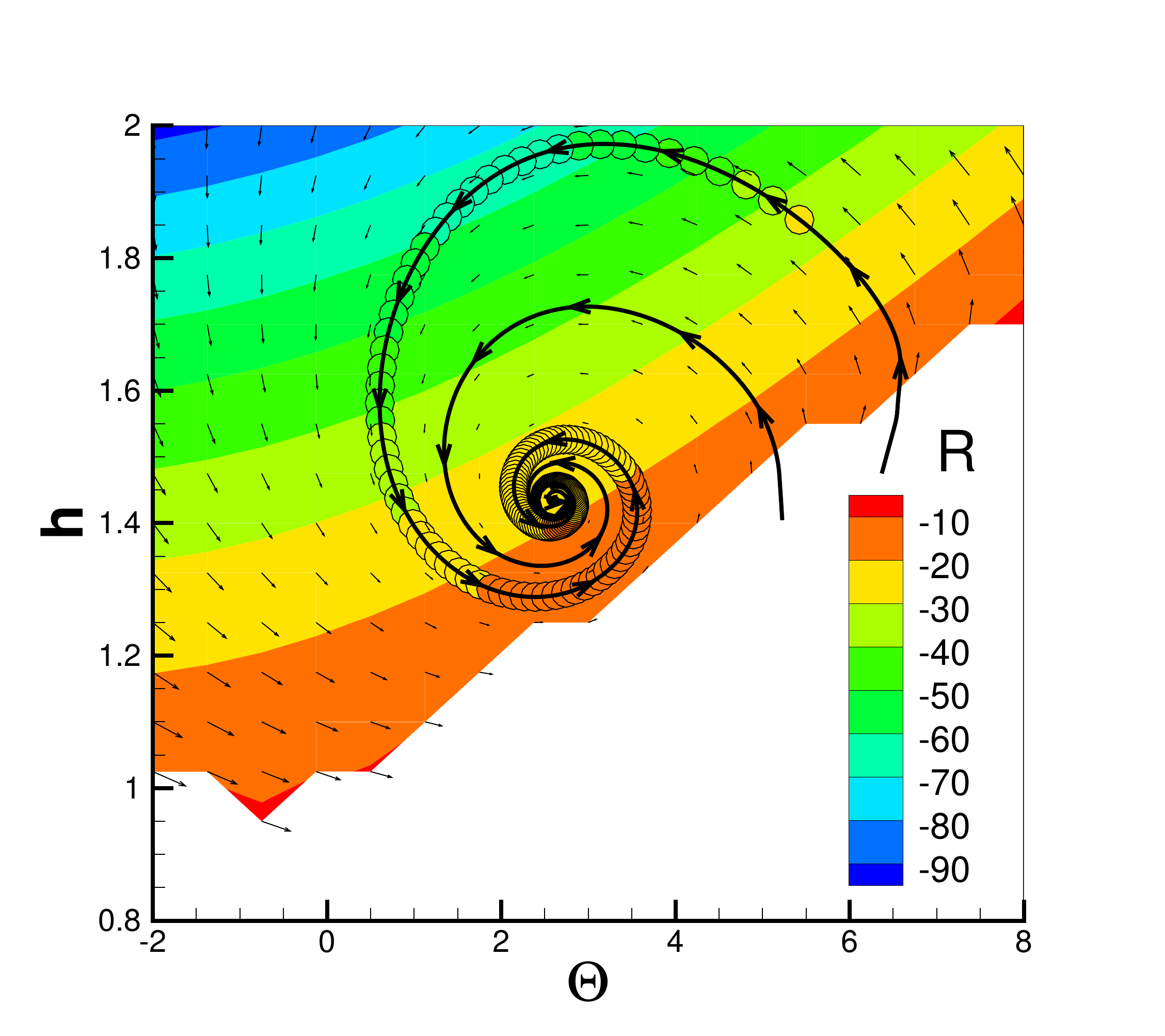}} 
  \subfigure[]{\includegraphics[width=0.495\textwidth]{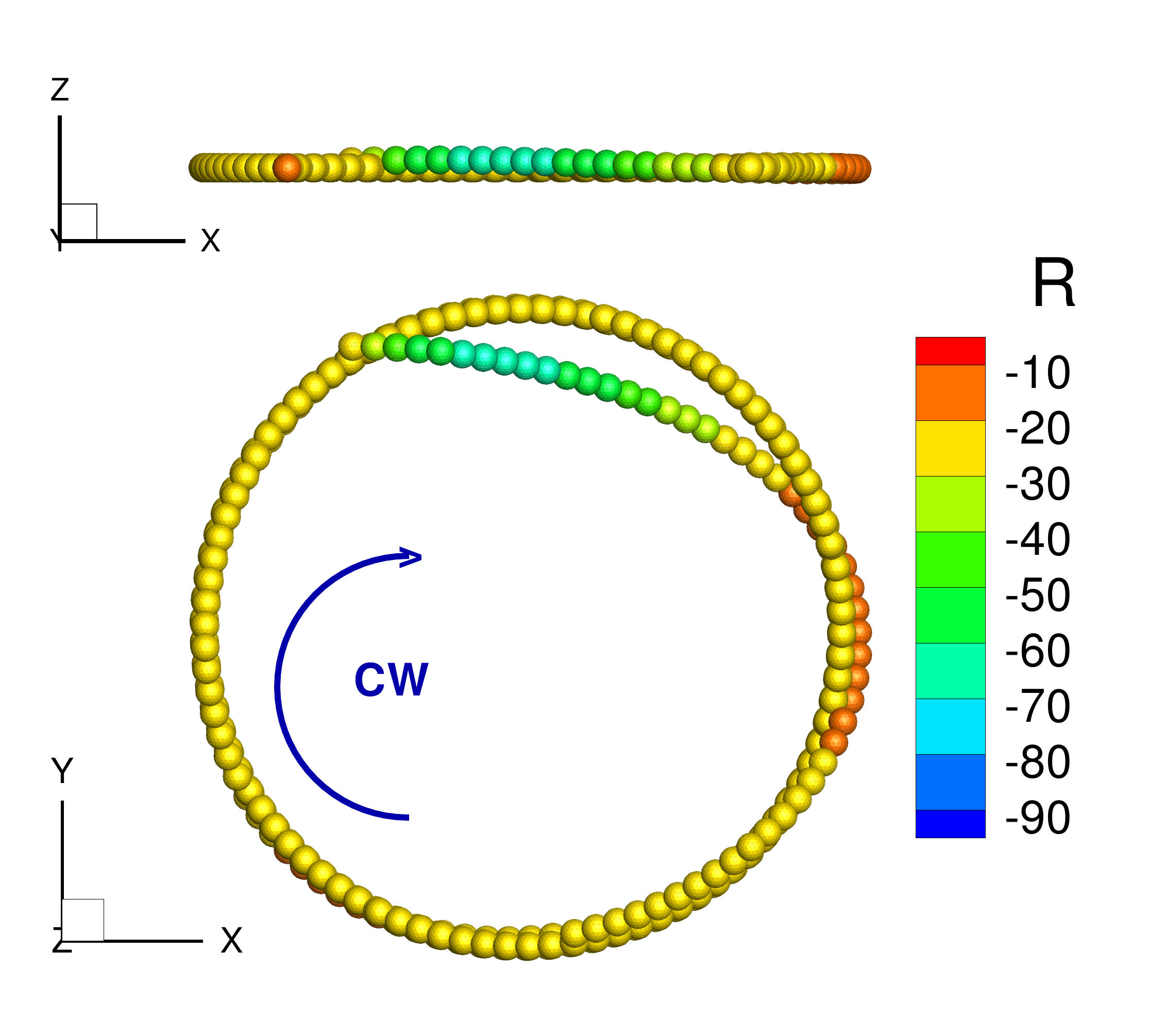}}
  \subfigure[]{\includegraphics[width=0.495\textwidth]{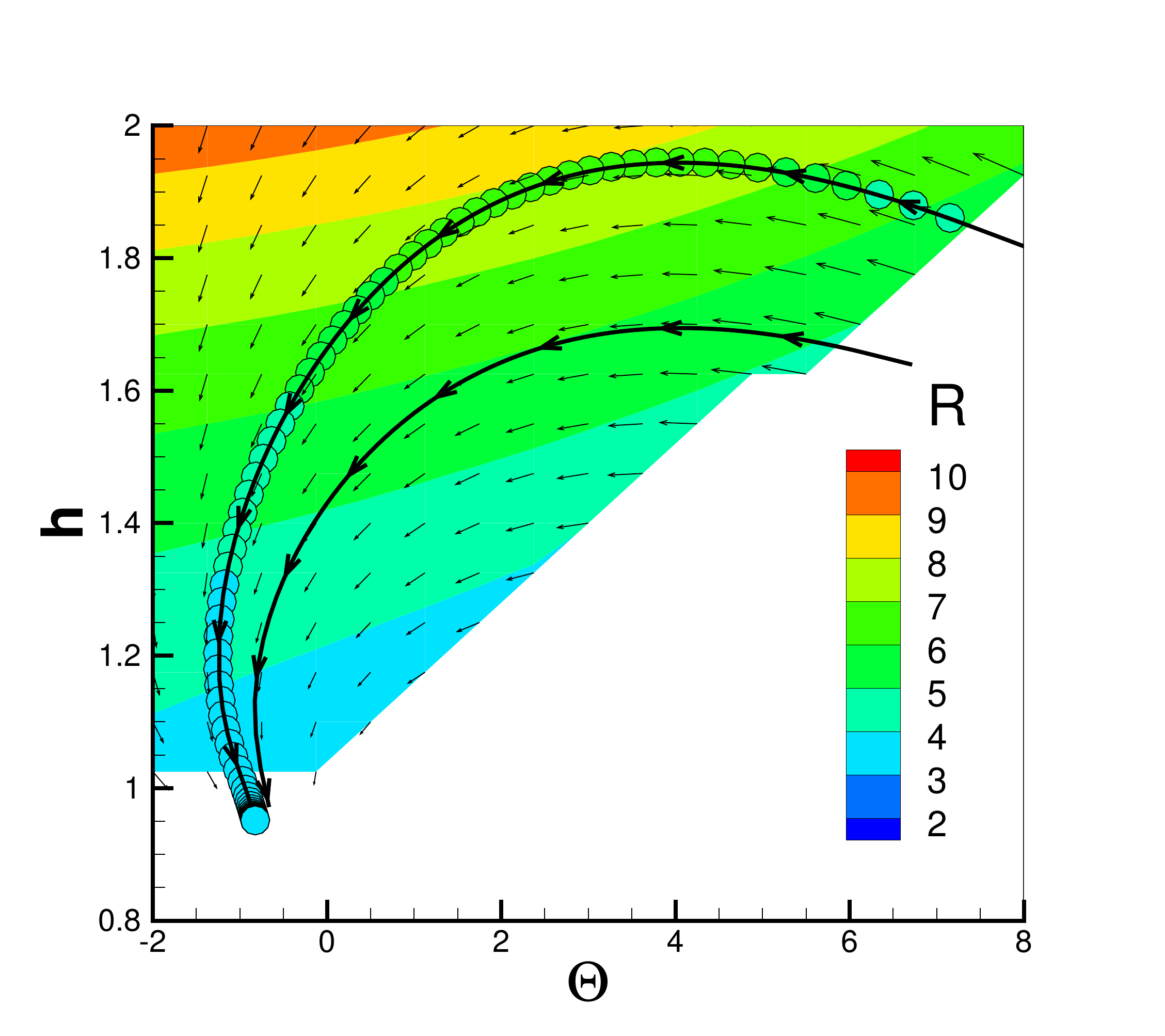}}
  \subfigure[]{\includegraphics[width=0.495\textwidth]{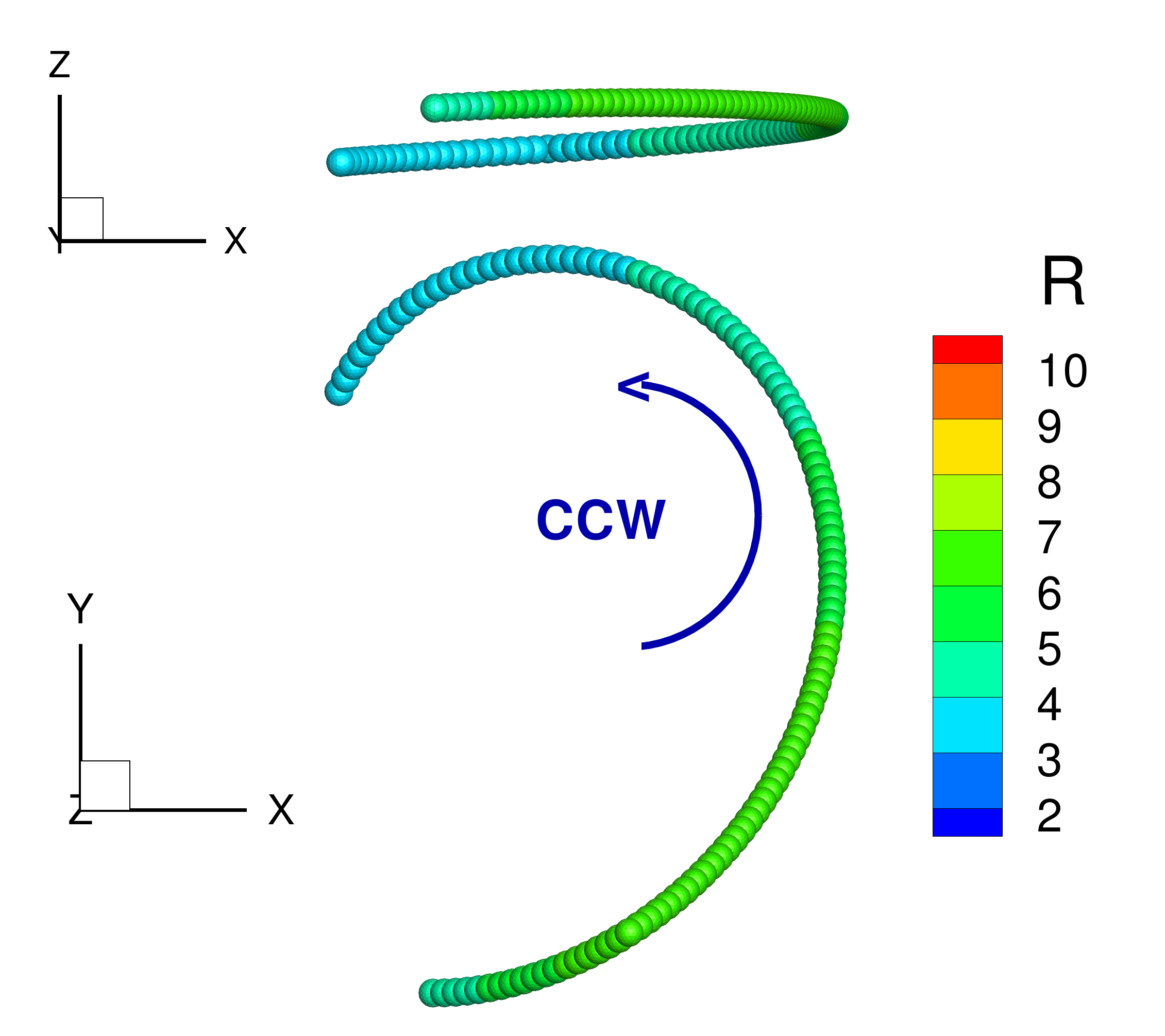}}
  \caption{
  Panel~(\textit{a}): phase plane $\hat{\Theta}-\hat{h}$ for a microswimmer
  close to a rigid no-slip wall. The curvature radius of the trajectory,
  equation (\ref{eqn:traj_radius}), is provided by the color map.
  Due to the presence of the wall the blanked regions are forbidden to
  the swimmer. Three typical trajectories $I$, $II$ and $III$ are
  shown by the solid black lines. Trajectories
  $I$ and $II$ converges to the stable point. The corresponding 
  trajectory in physical space is shown in panel~(\textit{b}) 
  where the curvature radius in the $XY-$plane is colour coded.
  Results for the free-slip interface are shown in
  panels (\textit{c}) and (\textit{d}). 
  All trajectories eventually intersect the planar interface 
  and no stable orbit exists. 
  Note the opposite sign of the curvature radius in comparison with the
  no-slip case.
  }
  \label{fig:NS_maptraj}
 \end{figure}

\subsection{Phase plane analysis and trajectories for a no-slip wall}
\label{ssec:PhField}

The interaction between the swimmer
and a homogeneous planar surface is completely determined by three parameters:  
the distance $h$ between the junction and the surface, the flagellum rotation 
phase angle $\phi$ and the pitch angle $\Theta$ between the swimmer longitudinal 
axis $\vec{e_1}$ and the surface plane, see figures~\ref{fig:geometry} 
and~\ref{fig:pitch_angle}. 
$\phi$ is a fast variable, thus, following~\cite{shum2010modelling}, 
the kinematics of the system can be described in term of 
$\phi -$averaged quantities, in the following denoted by a circumflex. 
The kinematics in the reduced $(\hat{\Theta},\hat{h})$
space is ruled by 
\begin{eqnarray}
\dot{\hat{\Theta}} &=& \hat{\Omega}_2 \,\, ,\\
\dot{\hat{h}} &=& \hat{U}_Z \,\, , 
\label{eqn:Theta_h_punto}
\end{eqnarray}
where $\hat{\Omega}_2 (\hat{\Theta},\hat{h})
= \hat{\vec \Omega}_H \cdot {\vec{e}_2}$ 
and $\hat{U}_Z(\hat{\Theta},\hat{h})= \hat{\vec{U}} \cdot {\vec{Z}}$. 
Figure~\ref{fig:NS_maptraj}(\textit{a}) shows the phase plane for a 
microswimmer close to a solid no-slip surface.
Three reduced trajectories, $I$, $II$ and $III$ 
are traced. The trajectory $III$ hits the wall while
trajectories $I$ and $II$, although starting from different initial 
configurations, both converge to the same attractor. 
The attractor is an 
asymptotically stable equilibrium point~\citep{cencini2009chaos}, 
($\hat{\Omega}_2$ and $\hat{U}_Z$ are zero), i.e. 
nearby trajectories converge to the equilibrium point.

The $\phi -$averaged trajectory in 3D space corresponding to 
$I$ is obtained by integrating 
the $\phi -$averaged version of the 
kinematic equations~(\ref{eqn:kinematic1}), (\ref{eqn:kinematic3}) 
and is reported in figure~\ref{fig:NS_maptraj}(\textit{b}). Apparently, 
after a transient, the swimmer stabilizes on a circular clockwise (CW)
orbit which corresponds to the  
stable point in the reduced space $(\hat{\Theta},\hat{h})$.
Following~\cite{shum2010modelling}, the curvature radius of the stable orbit 
is given by 
\begin{equation}
R = \frac{|\vec{\hat{U}(\hat{\Theta},\hat{h})}|}
{\hat{\Omega}_Z -\hat{\Omega}_X \tan{\hat{\Theta}}} \,\, ,
\label{eqn:traj_radius}
\end{equation}
which yields $R=-22.1$. As expected this result 
matches within numerical accuracy 
its direct measure $R=-21.9$ obtained from the 3D 
trajectory, see figure~\ref{fig:NS_maptraj}(\textit{b}). 
Note that equation (\ref{eqn:traj_radius}) differs in sign 
from the original equation given in~\cite{shum2010modelling}
due to a different choice of the body reference frame. It is also worth 
noting that equation (\ref{eqn:traj_radius}) provides the exact curvature radius 
only for the stable orbit. Nevertheless,
it also gives a reasonable approximation
in the other conditions here explored, as discussed below. 
In figure~\ref{fig:NS_maptraj}(\textit{a}) 
the color map refers to $R$ as estimated from eq.~(\ref{eqn:traj_radius}). 
The color code in figure~\ref{fig:NS_maptraj}(\textit{b}) 
corresponds to the local curvature radius of the trajectory 
projection in the $xy$-plane. This information is also reported along the curve 
$I$ in the phase plane, panel~\ref{fig:NS_maptraj}(\textit{a}), through the 
color of the open circles superimposed to the trajectory $I$. From the data it 
is apparent that the color differences between the circles
and the color map in the background can hardly be appreciated. 

\subsection{Swimming in presence of a free-slip interface}\label{ssec:free_slip}

The phase plane analysis has been performed for a microswimmer moving 
close to a free-slip interface, as illustrated in 
figures~\ref{fig:NS_maptraj}(\textit{c}) and~\ref{fig:NS_maptraj}(\textit{d}).
The first significant result is the sign of the curvature 
radius, now positive, i.e. the swimmer exhibits a counter-clockwise (CCW) motion 
in contrast to what observed for the no-slip wall, compare the panels (\textit{b}) 
and (\textit{d}) in figure~\ref{fig:NS_maptraj}.
The curvature radius is smaller with respect to the no-slip case
at corresponding $(\hat{\Theta}, \hat{h})$. 
The curvature radius measured in our simulation favorably
compares with experimental observation of~\cite{di2011swimming} where 
a CCW motion with a curvature radius of $\simeq 10 \mu m$ is described. 
Indeed, assuming $\bar{a}=1 \mu m$, appropriate for an E.coli, 
the typical radius of curvature found from the present numerics, $R \simeq 7$,
gives $R' = R \bar{a} \simeq 7\mu m$.
 
The second significant result 
is that no stable trajectory exists when the 
swimmer moves close to a free surface. In fact, a swimmer similar to an E.coli moving
almost parallel to the free-surface (low $\hat{\Theta}$) is always attracted to it. 
This result confirms the
indication of the 2D model proposed by~\cite{crowdy2011two}
for undeformable interfaces.
This conclusion also partially agrees with the results illustrated by~\cite{lopez2014dynamics}, 
where the authors, basing on a multipole expansion technique, highlighted that the average surface-normal 
velocity $\hat{U}_z$ of the swimmer is always negative for $\hat{\Theta} = 0$. Our work adds that the surface 
is no more able to attract the swimmer if it swims above a certain height $h$, even with 
zero tilt angle. 

\begin{figure}
  \includegraphics[width=0.47\textwidth]{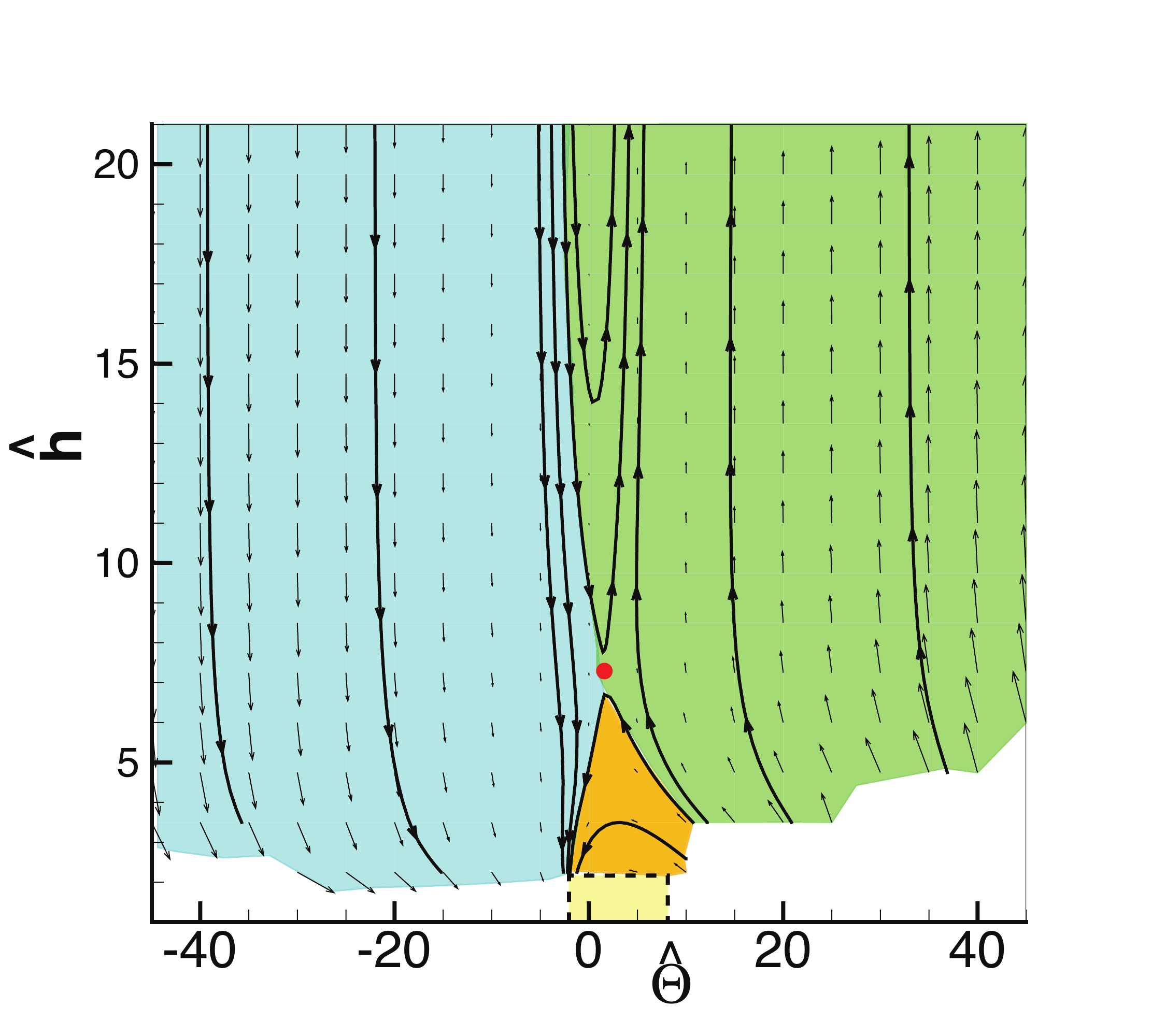}
  \includegraphics[width=0.47\textwidth]{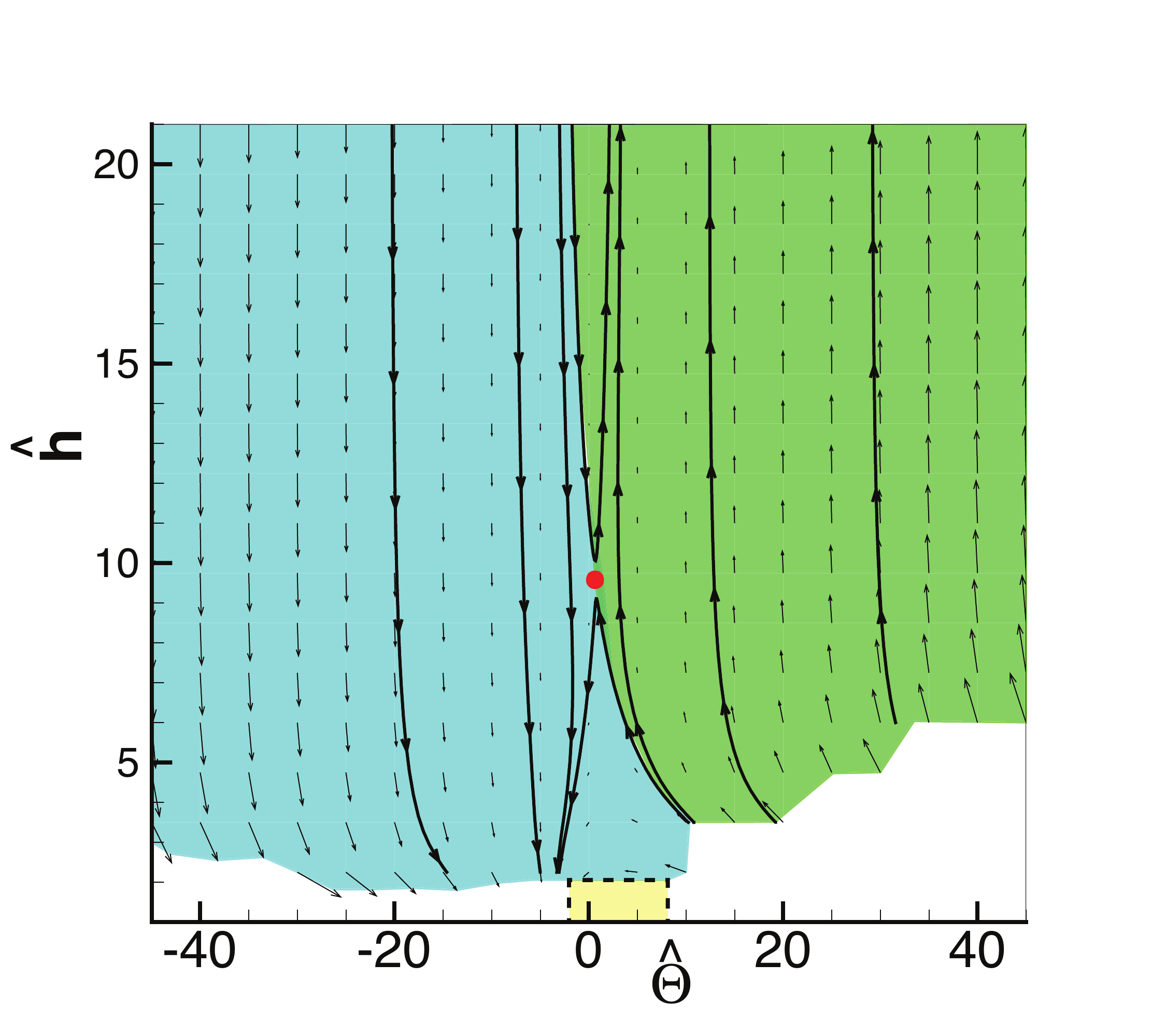}
  \caption{A wider view of the phase plane is displayed for
  the microswimmer near the no-slip (left) and the free-slip (right)
  surface. The range exploited in figure~\ref{fig:NS_maptraj} 
  correspond to the small dashed yellow box.
  Three regions of the phase plane are identified by colours:
  a trajectory belonging to the
  light blue region ends colliding with wall, one in the green
  region escapes away, and trajectories in the orange region are 
  attracted to the stable orbit. 
  Note that attraction basin is absent in the free-slip case (right).
  Unstable equilibrium points are marked by the red circles.
  }
\label{fig:large_angles}
\end{figure}

In figure ~\ref{fig:large_angles} we extend our analysis to a wider range 
of initial conditions, for both no-slip and free-slip surfaces. 
In both cases, at high positive $\hat{\Theta}$ the microswimmer escapes from the 
surface while, at high negative $\hat{\Theta}$, it eventually hits the boundary. 
On the no-slip wall three cases are possible depending on the initial 
conditions, namely the swimmer achieves a stable orbit, it escapes 
from the wall or it collides with the wall. 
In contrast, for a free-slip surface only 
two possibilities exist, namely, 
escaping from the surface or being attracted towards the interface.
The maps also highlight the presence of a further equilibrium point 
$(\hat{\Theta}_u,\hat{h}_u)$
 where the time derivatives 
of both ${\hat{h}}$ and $\hat{\Theta}$ vanish. However, as shown 
by the behavior of the streamlines in the neighbor of
$(\hat{\Theta}_u,\hat{h}_u)$, this 
point is unstable for both no-slip and free-slip cases, i.e. 
nearby trajectories escape from the equilibrium point.

\begin{figure}
  \includegraphics[width=\textwidth]{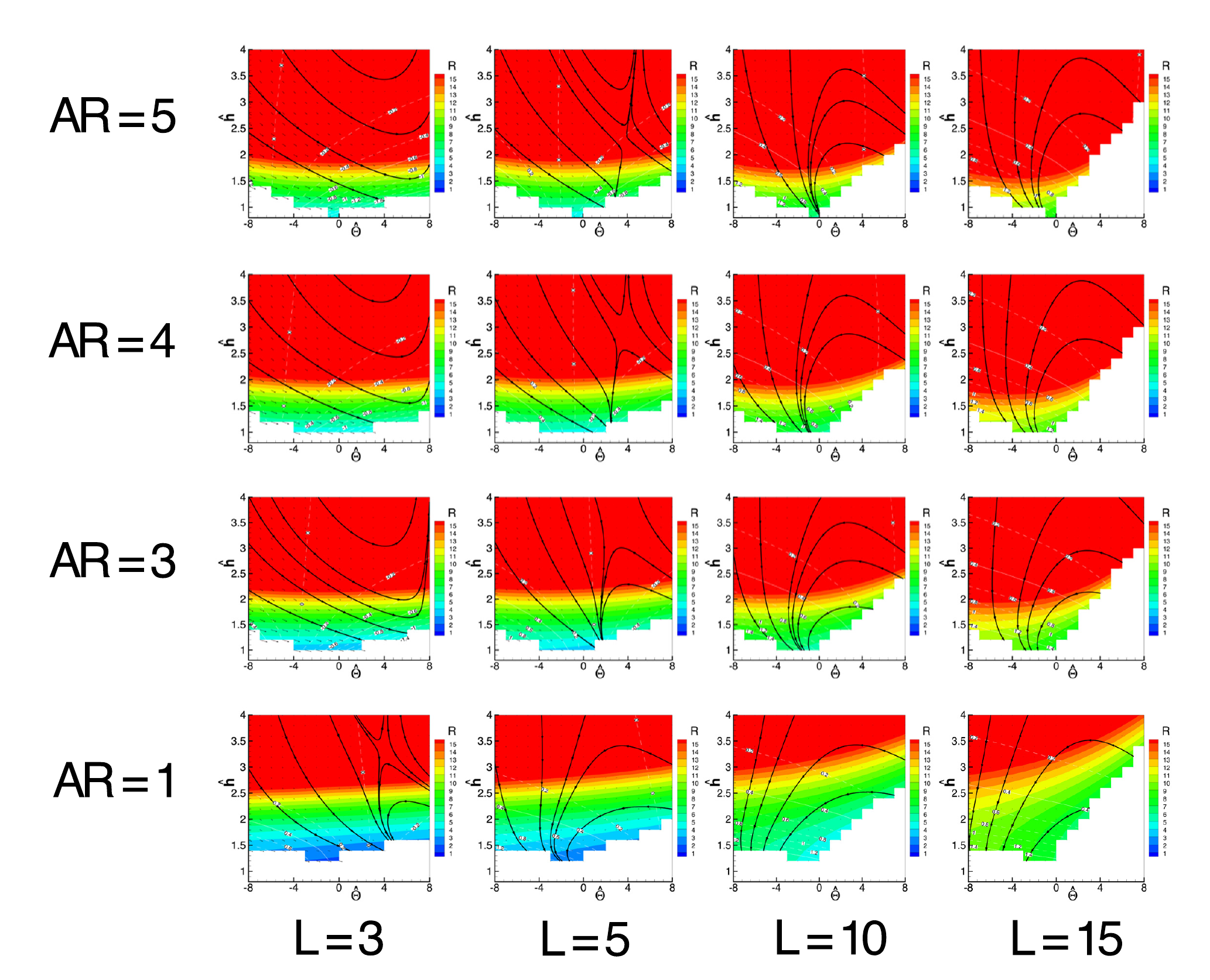}
  \caption{Phase plane $\hat{\Theta}-\hat{h}$ for a microswimmer
  close to a to a free-slip interface for different geometries, namely 
  head aspect ratio $AR=a_1/a_2 \in (1,3,4,5)$ and  tail length $L \in (3,5,10,15)$.
  The color map corresponds to the curvature radius along trajectories (e.g. solid black lines).
  }
\label{fig:many_geometries}
\end{figure}

In order to extend the analysis, the dynamics of swimmers with different geometrical 
characteristics was also examined, focusing in particular on the shape of the head  and the tail (axial) length. 
Figure~\ref{fig:many_geometries} shows the phase plane $\hat{\Theta}-\hat{h}$
when the tail length $L \in (3,5,10,15)$ and the head aspect ratio $AR=a_1/a_2 \in (1,3,4,5)$. 
When changing the aspect ratio, the swimmer does not substantially modify its behavior. In particular,  
the orientation as described by the pitch angle $\hat \Theta$ does not qualitatively change and the regions of phase space where
the swimmer is attracted to the free-surface or where it escapes away from it remain very similar. The general trend is an increase of the curvature of the trajectory for given $\left({\hat \Theta}, \, {\hat h}\right)$ as $AR$ increases.
On the other hand,  when changing  the relative tail-to-head length,  the swimmer hits the wall with different pitch angles 
$\hat \Theta$.  Decreasing the tail length, the region of phase plane where the swimmer escapes  becomes larger while the region where it is attracted  to
the free-surface shrinks.  The curvature of the trajectory, for  given $\left({\hat \Theta}, \, {\hat h}\right)$,  is found to increase with the tail length.

\subsection{Drift along the trajectory}\label{ssec:drifting}

\begin{figure} 
\centering
\subfigure[]{\includegraphics[width=0.9\textwidth]{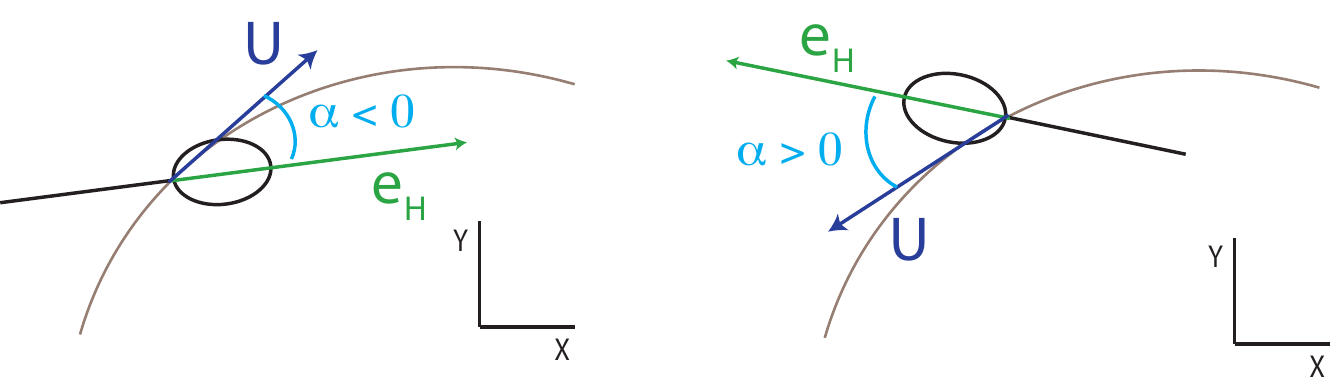}}
\subfigure[]{\includegraphics[width=0.49\textwidth]{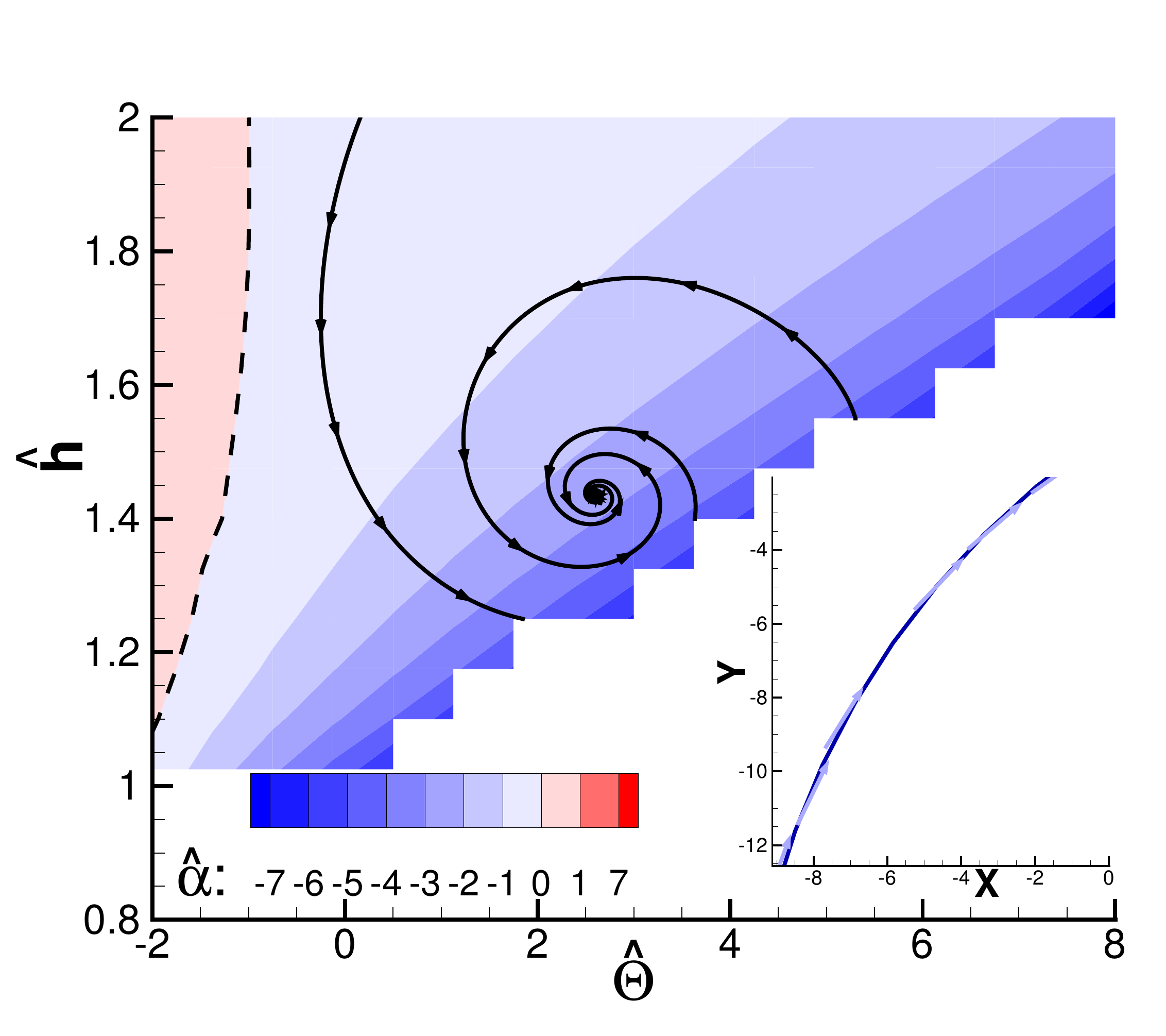}}
\subfigure[]{\includegraphics[width=0.49\textwidth]{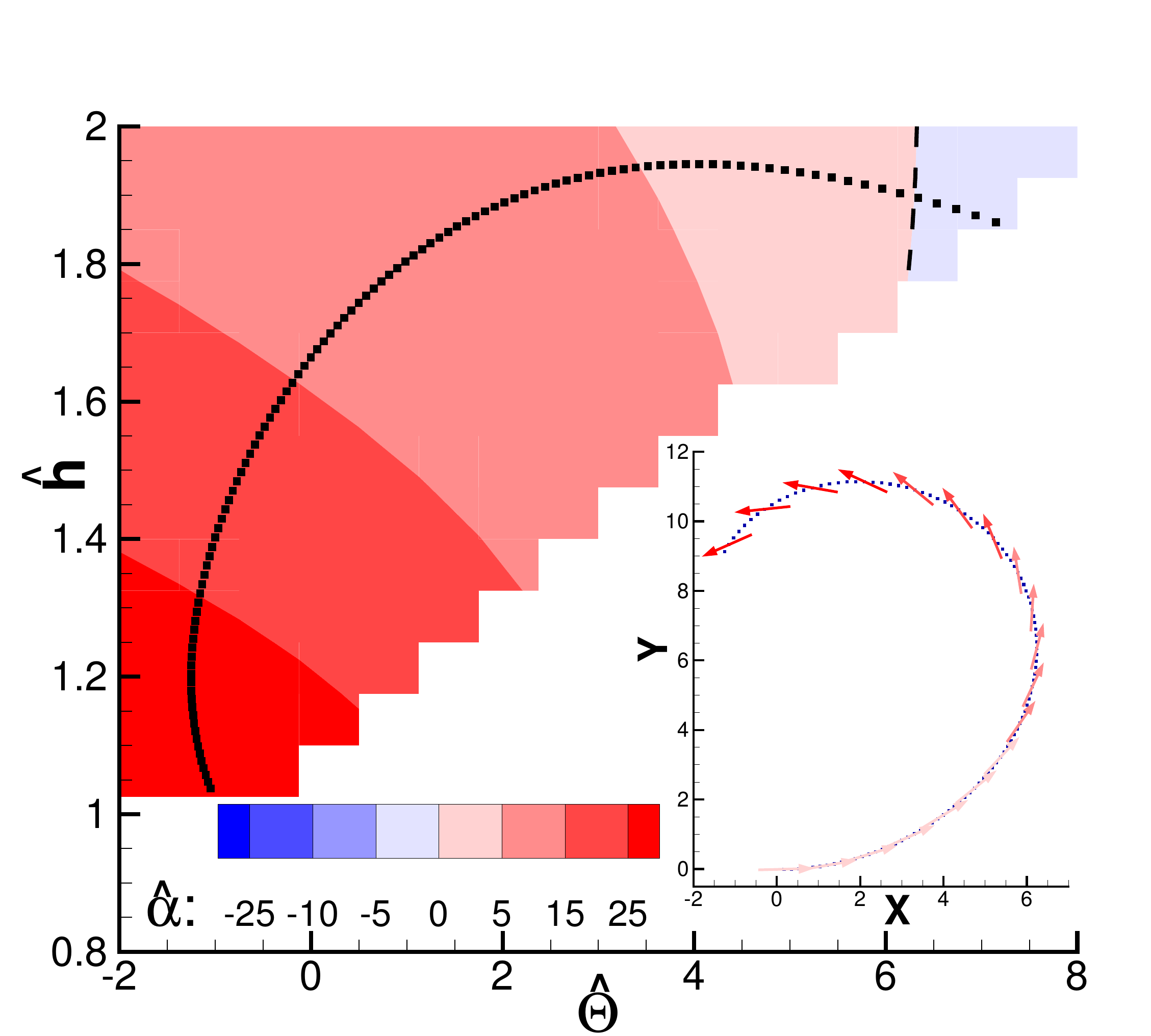}}
\caption{Panel (\textit{a}), sketch illustrating the drift angle
$\alpha$. Panels (\textit{b}) and (\textit{c}) report 
the $\phi-$averaged drift angle $\hat{\alpha}$ in the phase-plane (colour code),
for the no-slip and free-slip interface, respectively.
Each inset show a representative trajectory with superimposed the
longitudinal unit vector $\vec{e}_1$ (the color coding corresponds to the
local value of $\hat{\alpha}$).
}
\label{fig:drift_traj}
\end{figure}

Typically the longitudinal axis of  the microswimmer is misaligned with the local velocity evaluated at the junction point.
This behavior was experimentally observed for \textit{E.Coli}
moving close to a liquid-air interface by~\cite{di2011swimming}. 
In order to compare with experiments we define the drift angle $\alpha$ as 
the angle between the ${X,Y}$ projections of the body frame unit vector 
$\vec{e}_1$ and the swimmer velocity $\vec{U}$. 
The sign of $\alpha$ is taken to be positive when the swimmer points 
outward with respect to the trajectory, see figure~\ref{fig:drift_traj}a.

In figure \ref{fig:drift_traj}b and c, the $\phi-$averaged 
drift angle is reported for both no-slip and free-slip cases.
In the free-slip case the head points outside the trajectory
and $\hat{\alpha}$ increases as the swimmer approaches the wall.
The spanned range of values, $\hat{\alpha} \in (10^\circ,30^\circ)$, is 
in good agreement with the experimental observation~\citep{di2011swimming}.
In contrast, in the no-slip case, $\hat{\alpha}$ is slightly 
negative, $\hat{\alpha} \simeq -2^\circ$, meaning that the swimmer is
almost aligned with the trajectory of the reference point $\vec{x}_J$. 

\begin{figure} 
\centering
\includegraphics[width=\textwidth]{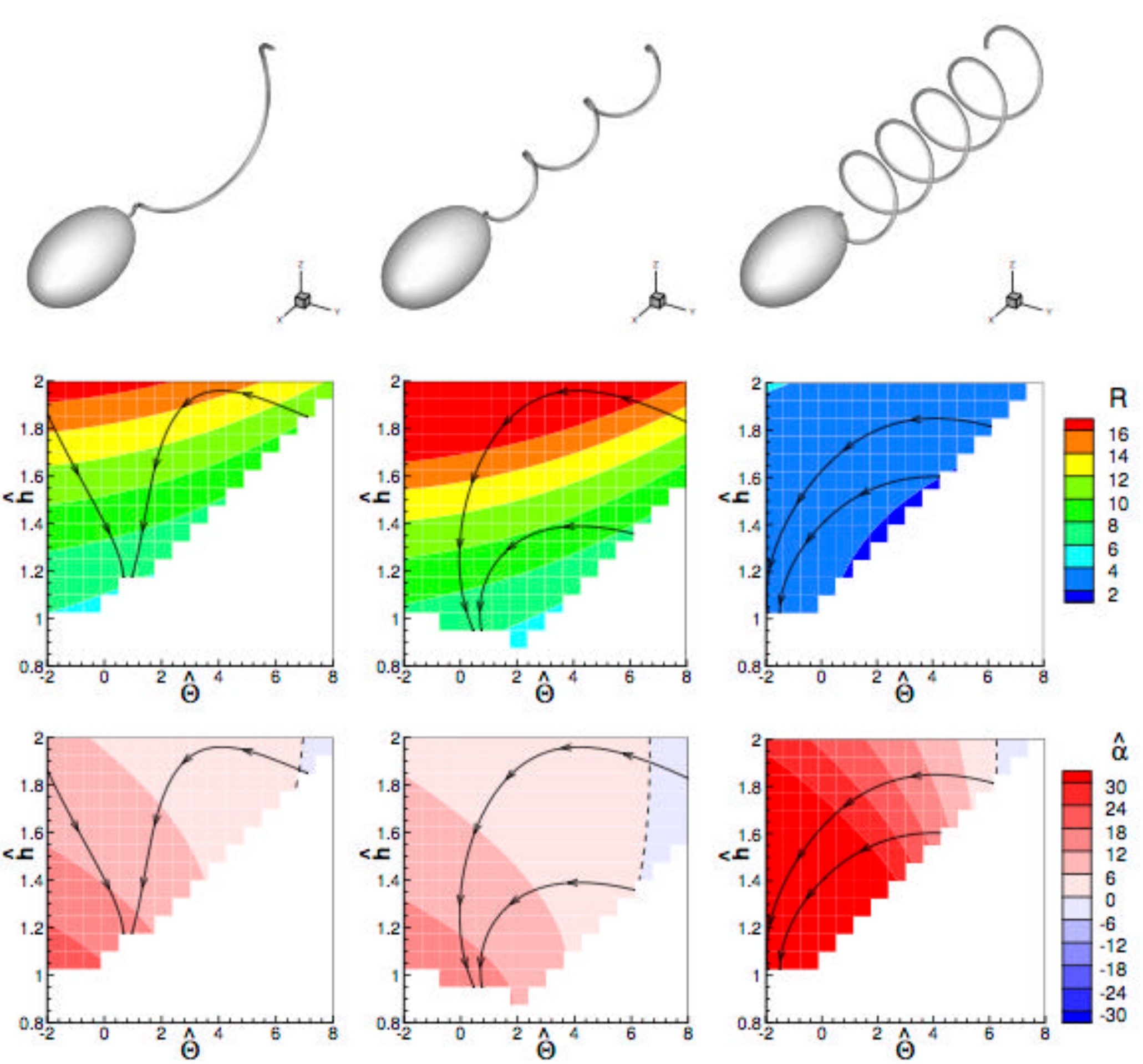}
\caption{Different tail geometries. This figure shows the radius of curvature $R$ and the 
drift angle $\hat\alpha$ when modifying the tail amplitude $A$ or the 
number of turns $N_\lambda$. Three cases are reported: 
$A=0.8$ and $N_\lambda=1$ (left), 
$A=0.4$ and $N_\lambda=3$ (center), 
$A=0.8$ and $N_\lambda=5$ (right). 
The hydrodynamic behavior is qualitative the same observed for the reference configuration
in figures~\ref{fig:NS_maptraj} and~\ref{fig:drift_traj} with 
few changes in the variables values.}
\label{fig:imm_grande}
\end{figure}

Here, to complete the discussion,  the effect of modifying the tail geometry is briefly addressed.
Indeed, among the large number of different parameters
defining the flagellated geometry, probably the most 
uncertain ones concern the tail, specifically amplitude of the helix, $A$, and
number of turns, $N_\lambda$.
Figure~\ref{fig:imm_grande} reports the radius of curvature $R$ and the drift angle
$\hat\alpha$ for three different tails. The comparison shows that the results discussed so far are generic. 

\section{Conclusions}\label{sec:conclusions}

The motion of a flagellated microswimmer close to a boundary, either a solid wall or 
a free surface, is relevant to several applications spanning from micro-robots to biology and medicine, as concerning in particular biofilm formation. Its dynamics can in principle be affected by several physical phenomena occurring at 
the microscale and, for biological applications, by the behavior of the microorganism.
The chemical and physical nature of the interface may play a role in the interaction between the
swimmer and the surface, e.g. surface charges and chemicals adsorbed at the interface may have a significant influence. 
Concerning the specific case of E. coli, taken as representative of most flagellated, recent experimental studies reported a characteristic motion
of the swimmer in the two extreme cases of a solid wall and a free surface.  In the first case, the experimental observation
consistently show that the microswimmer typically moves in circulatory orbits oriented in clockwise direction (CW)  \cite{lauga2006swimming}.
In contrast, there is evidence that the orientation of the trajectory is reversed (counterclockwise, CCW) when swimming occurs near a free surface \cite{di2011swimming}.
However, the effect of the free surface is less neatly defined and, sometimes, CW motion is reported, probably due to
the presence of contaminants adsorbed at the interface  \citep{lemelle2010counterclockwise}.
Theoretical models proposed to explain this behavior typically exploit several form of approximation, e.g. multipole expansion, that may become
partially inaccurate when the microswimmer gets very close to the interface.

The present paper provides a complete description of the motion, considering a reasonably realistic geometry of the flagellated in 
presence of a  free-surface modeled as a rigid, free-slip plane.
The resulting model removes any concurrent effect, retaining a full hydrodynamics description.
The results for the free-slip boundary were compared with those already analyzed in~\cite{shum2010modelling} for the no-slip case.
The  data clearly indicate that the motion close to a  liquid-air interface is CCW, in agreement with  the experimental data~\citep{di2011swimming}
and with the theoretical results  obtained by using multipole expansions~\citep{lopez2014dynamics} and resistive force theory~\citep{di2011swimming}.

Other available experimental information, namely the orientation of the bacteria with respect  to its trajectory~\citep{di2011swimming},
is satisfactorily reproduced by the present simulations, confirming that the head of the swimmer points outward.
In contrast, the bacteria is roughly aligned with its trajectory close to a no-slip wall.
In principle,  this observable can be used together with the rotation direction to interpret experimental results on the interaction of a 
microswimmer with an interface.
A characteristic aspect of the motion near a solid surface is the occurrence of stable orbits \citep{giacche2010hydrodynamic,shum2010modelling}.
To the contrary, no stable orbit has been presently found on a free-slip interface. 

In conclusion, the boundary conditions are confirmed to deeply influence the hydrodynamical behavior of the swimmer.
This consideration paves the way to suitably textured surfaces which, properly engineered  to stably support a 
super-hydrophobic state, can be exploited to passively control the microswimmer motion. 
Indeed the ability of superhydrophobic surfaces to alter the flow has been recently used for passive particle separation
\citep{asmolov2015principles}. Under this respect fully resolved hydrodynamic simulations able to model complex physical surfaces, 
like the BEM adopted here, could provide the required fundamental knowledge to extend passive control strategies to active suspensions.
\\

The authors acknowledge the CINECA Iscra C Award (IscrC-BSM-LAI) for the availability of HPC resources
and the ERC Grant  No. [339446] {\bf BIC}: {\sl Bubbles from Inception to Collapse}.

\appendix
\section{Green functions for a flat free-slip interface}\label{appA}
\begin{figure} 
\centering
\includegraphics[width=0.32\textwidth]{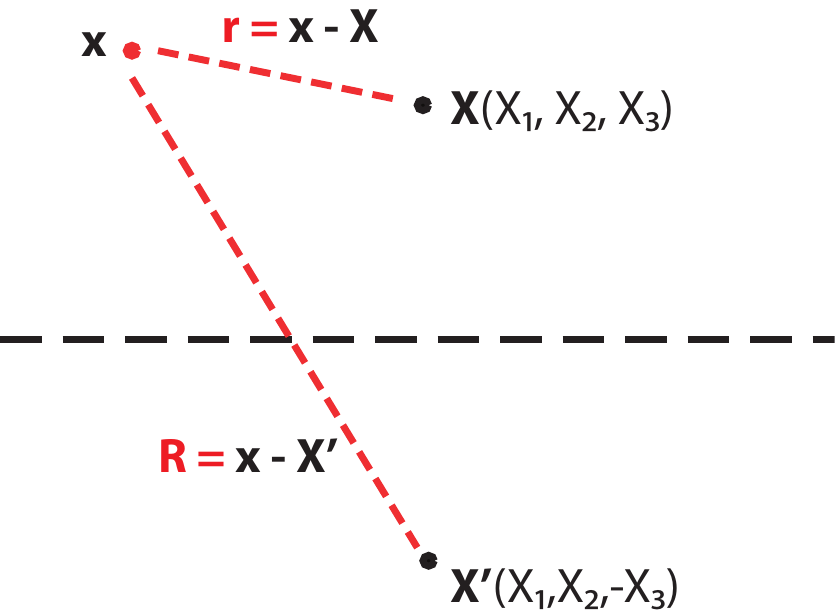}
\includegraphics[width=0.32\textwidth]{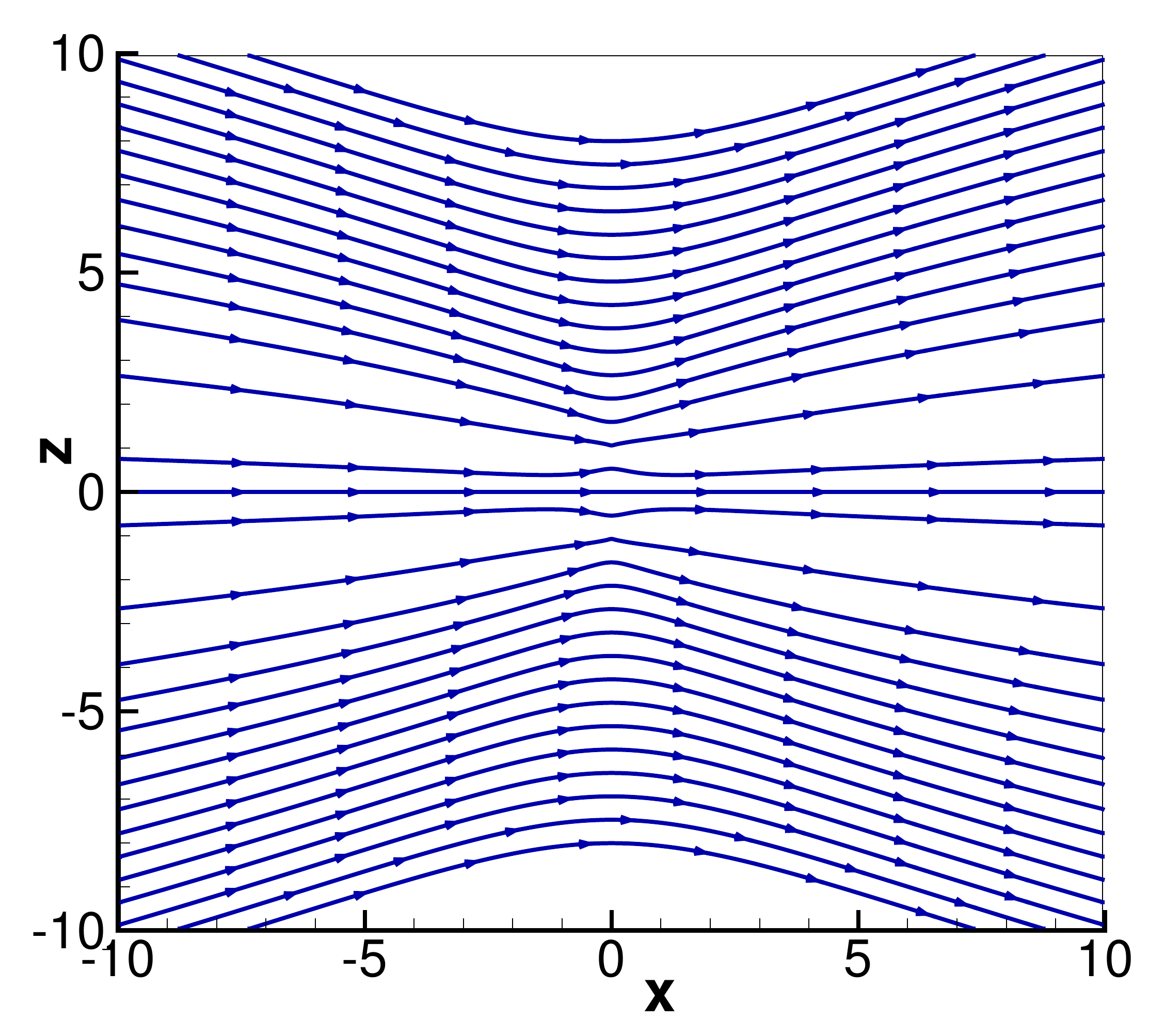}
\includegraphics[width=0.32\textwidth]{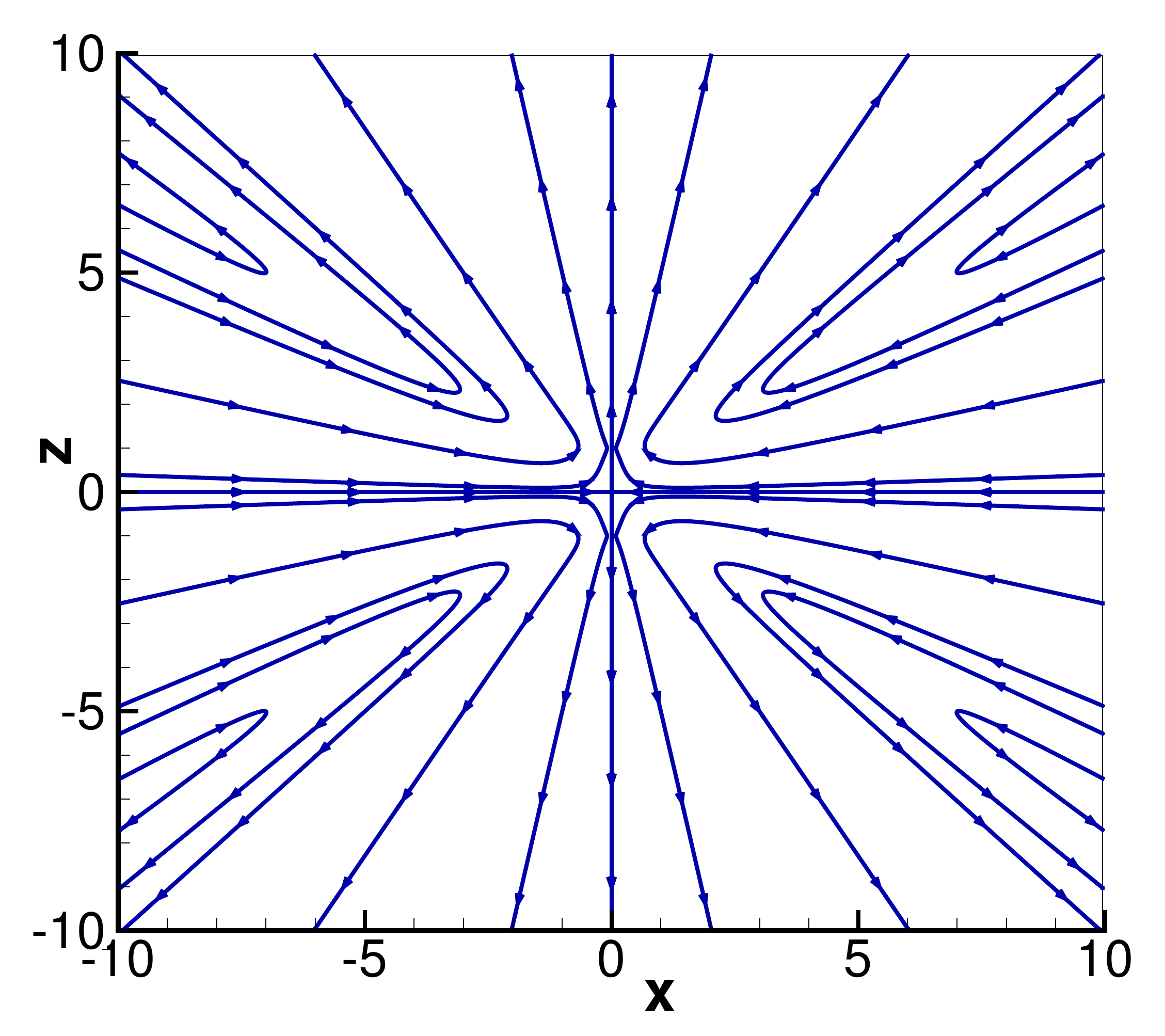}
\caption{Left) The image system employed to obtain the Green function 
for a planar free-slip interface. 
Center and right panels report the 
streamlines in the $x_2=0$ plane, due to a point force $\vec{b}=(1,0,0)$
in $\vec{X}=(0,0,1)$ and a point force $\vec{b}=(0,0,1)$ in $\vec{X}=(0,0,1)$,
respectively. 
In both cases the velocity at the interface is purely parallel 
to the $x_3=0$ plane, as required by the impermeability. 
The symmetry of the fields with respect to $x_3 = 0$ implies the
vanishing of the tangential shear stress at the interface.
\label{fig:stream_x}
}
\end{figure}

The Green's function for a solid adherent wall is known
since the work of~\cite{blake1971note}. 
The corresponding fundamental solution for a free-slip interface does not seem to be mentioned in the
current literature. As shown here, it can be obtained
by means of the method of images where the generic localized 
force is reflected with respect to the plane in such a way 
to satisfy the impermeability and the zero tangential stress condition. 
For convenience the index notation is adopted here
with $\vec{x}=(x_1,x_2,x_3)$ the coordinates of a point in space. 
The image system is reported in 
figure~\ref{fig:stream_x} (left panel) where the point force
$\vec{b} =(b_1,b_2,b_3) $ placed at $\vec{X}=(X_1,X_2,X_3)$ 
is reflected in $\vec{X'}=(X_1,X_2,-X_3)$ as
$\vec{b'} =(b_1,b_2,-b_3)$.
The relative positions
\begin{eqnarray}
\vec{r}  & = &\vec{x}-\vec{X}=(x_1-X_1,x_2-X_2,x_3-X_3) \\
\vec{r}' & = &\vec{x}-\vec{X'}=(x_1-X_1,x_2-X_2,x_3+X_3) 
\end{eqnarray}
connect the generic point $\vec{x}$ in the fluid domain
with the two singularities in $\vec{X}$ and $\vec{X}'$, respectively.
Taking into account the well known expression for the free space
Green function~\citep{happel1983low}
\begin{equation}
G^{free}_{ij}=\left( \frac{\delta_{ij}}{r}+\frac{r_i r_j}{r^3} \right) \, ,
\end{equation}
the superposition of the two singularities yields
\begin{eqnarray}
G_{i1}^{FS}=\left( \frac{\delta_{i1}}{r}+\frac{r_i r_1}{r^3} \right)+ 
            \left( \frac{\delta_{i1}}{r'}+\frac{r'_i r'_1}{r'^3} \right) \, , \\
G_{i2}^{FS}=\left( \frac{\delta_{i2}}{r}+\frac{r_i r_2}{r^3} \right)+ 
            \left(\frac{\delta_{i2}}{r'}+\frac{r'_i r'_2}{r'^3} \right) \, , \\
G_{i3}^{FS}=\left( \frac{\delta_{i3}}{r}+\frac{r_i r_3}{r^3} \right)- 
            \left(\frac{\delta_{i3}}{r'}+\frac{r'_i r'_3}{r'^3} \right) \, .
\label{eqn:greenFS_III}
\end{eqnarray}
The i-th velocity component at $\vec{x}$ is expressed 
in terms of the force intensity $\vec{b}$ at $\vec{X}$ as
\begin{equation}
u_i(\vec{x})=\frac{1}{8\pi \mu} G_{ij}^{FS}(\vec{x}, \vec{X}) b_j \, .
\label{eqn:zz}
\end{equation}
It can be immediately checked that the ensuing velocity field
satisfies impermeability and free-slip conditions at $x_3=0$.
Indeed, as concerning the wall normal velocity, impermeability 
straighforwardly follows from (\ref{eqn:zz}).
Let's consider now the zero tangential stress condition.
From the stress tensor associated with (\ref{eqn:zz})
\begin{eqnarray}
T_{ik1}^{FS}=-6\frac{r_i r_1 r_k}{r^5}- 6\frac{r'_i r'_1 r'_k}{r'^5} \, , \\
T_{ik2}^{FS}=-6\frac{r_i r_2 r_k}{r^5}- 6\frac{r'_i r'_2 r'_k}{r'} \, , \\
T_{ik3}^{FS}=-6\frac{r_i r_3 r_k}{r^5}+ 6\frac{r'_i r'_3 r'_k}{r'} \, ,
\end{eqnarray}
one immediately realizes that the tangent component
of the tension at $x_3 = 0$ vanishes identically, 
i.e. $T_{i3j}^{FS} b_j = 0$ for $i = 1,2$. 
An illustration is provided in the two rightmost 
panels of figure~\ref{fig:stream_x} 
where the streamlines associated with free-slip Green function 
due to a point force respectively parallel $\vec{b}=(1,0,0)$ 
and normal $\vec{b}=(0,0,1)$ to the wall placed at $X=(0,0,1)$ 
are reported. 

\bibliographystyle{plain}

\bibliography{Pimponi_microswimmer}

\end{document}